\documentclass{epl}

\usepackage{amsmath,amsfonts,amssymb}

\renewcommand{\section}[1]{{\em #1}~-}
\newcommand{\beq}{\begin{equation}} 
\newcommand{\eeq}{\end{equation}}
\newcommand{\bqa}{\begin{eqnarray}} 
\newcommand{\eqa}{\end{eqnarray}}
\newcommand{\nn}{\nonumber}

\newcommand{\Tr}{\mathrm{Tr}}

\newcommand{\erf}[1]{Eq.~(\ref{#1})}

\newcommand{\dg}{^\dagger}

\title{Classical Robustness of Quantum Unravellings}
\author{D.J. Atkins\inst{1,2} \and Z. Brady\inst{1} \and K. Jacobs\inst{1,2} \and H.M. Wiseman\inst{1,2}}
\institute{
  \inst{1} Centre for Quantum Dynamics, School of Science, Griffith University,
  Brisbane, 4111 Queens\-land, Australia.\\
  \inst{2} Centre for Quantum Computer Technology, Australia.}

\pacs{03.65.Yz}{Decoherence; open systems; quantum statistical methods}
\pacs{42.50.Lc}{Quantum fluctuations, quantum noise, and quantum jumps}
\pacs{05.40.Jc}{Brownian motion}

\begin{document}

\maketitle

\begin{abstract}
We introduce three measures which quantify the degree to which quantum systems
possess the robustness exhibited by classical systems when subjected to
continuous observation. Using these we show that for a fixed environmental
interaction the level of robustness depends on the measurement strategy, or
unravelling, and that no single strategy is maximally robust in all ways.
\end{abstract}

When describing the observation of individual quantum systems, it is crucial to
explicitly treat the measurement process using quantum measurement theory. The
reverse, however, is commonly true for classical systems --- it is quite usual for
experimentalists both to model and gather data about classical systems without
any reference to classical measurement theory (being the theory of Bayesian
statistical inference or nonlinear filtering~\cite{BayesBook,Maybeck}). This dramatic
contrast is only possible because of certain properties which many classical
systems possess. Three such properties are apparent if we consider watching the
motion of a pendulum: 1. When we open our eyes, we obtain the information (the
location and velocity of the pendulum) almost instantaneously, 2. the
information we obtain is not exclusive --- many people can observe the same
system, and will all agree upon the results, and 3. the system is relatively
unaffected by noise so that if we close our eyes for a moment, we can
accurately predict what we will see when we open them again~\footnote{Note that
the existence of chaos in a system does not destroy its robustness: in this
case the  divergence of neighboring trajectories is exponential and thus has a
negligible effect on the predictability for short times.}. We will refer to the
degree to which a system possesses these properties as its degree of {\em
robustness} under each property. 

The classical robustness of quantum systems in this sense was the subject of a
recent study by Dalvit, Dziamarga and Zurek \cite{DalDziZur01}. They concluded
that, for a given environmental interaction, a single measurement strategy will
maximize the robustness for both properties 1 and 3 above. Here we investigate
the degree of robustness of two canonical quantum systems for the three
properties above (and a fourth), under a wide class of measurement strategies.
We show that this conclusion is, in fact, not warranted. That is,  in general,
for a fixed environmental  interaction, different measurement strategies are
required to maximize different notions of classical behavior, and that this
depends not only on the system in question, but the dynamical regime. There is
thus necessarily a trade-off between different types of robustness. While this
study is of fundamental interest, it is also of practical importance for the
feedback control of quantum systems~\cite{Belavkin,Doh00,WisManWan02}, as will
be discussed.

To clarify the above and establish our results, it is useful first to 
establish some concepts and terminology.  The continuous observation of a
quantum system can always be described by treating the interaction of the
system in question with an environment consisting of a large number of degrees
of freedom. As a result of the interaction, the environment continuously
extracts information about the system, and this information can be read by
measuring the state of the environment. This has been called quantum
filtering~\cite{Belavkin88}. Once one has chosen an interaction, one therefore
has the remaining freedom to choose the manner in which to interrogate the
environment, and different methods lead to qualitatively different kinds of
measurements. These different measurements, which constitute the measurement 
strategies discussed above, are often referred to as different {\em
unravellings} of the environmental interaction~\cite{Carmichael}.

In what follows, by {\em conditional evolution}, or {\em trajectory}, we will
mean the evolution of the system as a result of a particular unravelling
($\mathcal{U}$) and by {\em unconditional evolution} we will mean the evolution
of the system under the same environmental interaction, but without the
measurement of the environment. The unconditional evolution is simply given by
the ensemble average ({\it E}[$ \cdot$]) over all the possible conditioned
evolutions.

Before introducing quantitative measures of robustness which characterize the
speed, non-exclusivity and predictability of a system,  we briefly revisit the
conclusion of Ref.~\cite{DalDziZur01}. Using our terminology, they maintained
that the unravelling that most rapidly purifies the state is identical to the
one that will tend to collapse the system into a ``pointer state" \footnote{We
note that a ``pointer state'' unravelling similar to this was previously
considered in Ref.~\cite{DioKie00}.}. The ``pointer states"~\cite{ZurHabPaz93}
are those that lose their purity most slowly when the environment is
unobserved. If true, this would imply that the same unravelling is most robust
under our properties 1 and 3. Ref.~\cite{DalDziZur01} also argued  that purity
loss and fidelity loss were equivalent for the purposes of defining robustness.
We show that in general, contrary to {\em both} of the above conclusions, {\em
different} unravellings are optimal for these various concepts of robustness.

\section{Measures of robustness of unravellings}
We will consider four quantitative measures of robustness. The first, {\em
purification time}, quantifies the rate at which the measurement provides
information about the system.
To define this we allow the system to evolve to its unconditional steady state,
and then ask how long it takes, upon switching on the observation, for the
measurement to increase the purity halfway to its long-time value (which, for
an efficient measurement is unit purity). Formally, the purification time,
$\tau_{\mbox{\scriptsize pur}}$, is the earliest solution of
\begin{equation}
   E\left[
   \Tr\left[\left\{\mathcal{S}^\mathcal{U}_1(\tau,0)\rho_\mathrm{ss}\right\}^2\right]
   \right]=\theta ,
   \label{defPur}
\end{equation}
where $\theta\equiv\mbox{$\frac{1}{2}$}(1+\Tr[\rho_\mathrm{ss}^2])$. Here
$\rho_\mathrm{ss}$ is the unconditional steady state of the system and
$\mathcal{S}^\mathcal{U}_\eta(t_2,t_1)$ is the non-linear, stochastic,
completely positive map which takes the system from time $t_1$ to $t_2$ during
a measurement of efficiency $\eta$. It has the properties
\begin{equation}
  E\left[\mathcal{S}^\mathcal{U}_\eta (t,0) \right] \equiv \mathcal{O}(t,0) =
             \mathcal{S}^\mathcal{U}_0(t,0) = e^{\mathcal{L}t},
\end{equation}
where $\mathcal{L}$ is the Lindblad-type superoperator generating the
unconditional evolution.

The purification time is relevant for the feedback control of
quantum systems. Feedback control is realized by observing a system and using
this information, as it is obtained, to control the evolution (usually by
modifying the system Hamiltonian). The effectiveness of any feedback algorithm
is thus ultimately limited by the extent to which the system state is known.
The purification time therefore provides an indication of the time required for
a feedback loop to become effective, when it employs the given measurement
strategy.



Secondly, we wish to quantify the extent to which the existence of multiple
observers interferes with the ability of each observer to track the system. In
order for multiple parties to observe the system separately, they must divide
up the environment between them. This looks, to one of the observers, $i$, as
if she has an inefficient measurement, with efficiency coefficient $\eta_i$,
such that $\sum_i \eta_i = 1$ \cite{Bar93,DalZur04}. We will quantify the
robustness of the measurement scheme against a division of the environment by
asking at what value of efficiency the average purity of the long-time observed
state is halfway between its value at perfect efficiency, and that of no
observation at all. We will refer to this as the {\em efficiency threshold},
and denote it by $\eta_{\mathrm{thr}}$; formally this is the value of $\eta$
such that
\begin{equation}
   \lim_{t\rightarrow\infty} E\left[ \Tr\left[ \left\{ \mathcal{S}^\mathcal{U}
   _\eta(t,0) \rho_\mathrm{ss}\right\}^2\right]\right]=\theta .
   \label{defTh}
\end{equation}
The smaller $\eta_{\rm thr}$ is,  the more robust a feedback algorithm would
be in the face of inefficiency.

Our third measure of classical robustness, {\em mixing time}, quantifies the
rate at which the system becomes unpredictable --- essentially it characterizes
the sensitivity of the system to noise from its environment. To define the
mixing time, we allow the observed trajectory to evolve for long enough that
the state is pure, and that the unconditioned evolution would have reached a
steady-state, stop measuring the environment, and consider the decrease in the
purity of the state as time passes. The mixing time is the time at which the
purity falls halfway from its initial value (being unity) to the value it would
take if the system were allowed to evolve back to its steady state (being
Tr[$\rho_\mathrm{ss}^2$]). Formally, the mixing time, $\tau_{\mbox{\scriptsize
mix}}$, is the earliest solution of
\begin{equation}
   \lim_{t\rightarrow\infty} E\left[\Tr\left[\left\{\mathcal{O}(t+\tau,t)
                 \mathcal{S}^\mathcal{U}_1(t,0)\rho_\mathrm{ss}\right\}^2\right]\right] =
                 \theta.
   \label{defMix}
\end{equation}
This quantifies the predictability of the evolution while the observer
is not looking.

A related concept, introduced previously by two of us and
Vaccaro~\cite{WisVac,WisBra00}, is the extent to which an initially
conditioned state remains unchanged during a period of unconditional evolution.
This is quantified by the {\em survival time}, $\tau_{\mbox{\scriptsize sur}}$,
the earliest solution of \begin{equation}
   \lim_{t\rightarrow\infty} E\left[\Tr\left[ \mathcal{S}_1^\mathcal{U} (t,0)
   \rho_\mathrm{ss}\times\mathcal{O}(t+\tau,t)\mathcal{S}^\mathcal{U}_1(t,0)\rho_\mathrm{ss}\right]
   \right]=\theta .
  \label{defSur}
\end{equation} Note that the stochastic map $\mathcal{S}_1^\mathcal{U} (t,0)$
appearing twice in this equation is the same map (i.e. has the same noise). We
also note that $\tau_{\mbox{\scriptsize sur}}\leq \tau_{\mbox{\scriptsize
mix}}$. In general this does not measure the noise-sensitivity alone, since any
deterministic evolution will contribute to changing the state. While the
survival time is not motivated by classicality like the other measures of
robustness, it is, like them, relevant for feedback control. Consider feedback
with a time delay $\tau$. For a feedback algorithm designed for $\tau=0$, the
performance should not be greatly degraded as long as  $\tau \lesssim \tau_{\rm
sur}$, as the system will not have strayed significantly during the delay. On
the other hand, deterministic evolution during the delay could be corrected for
in the algorithm design. In that case, the performance of the feedback control
should not be significantly worse than for the case of no delay as long as
$\tau \lesssim \tau_{\rm mix}$.

\section{A particle undergoing quantum Brownian motion}
We consider a particle in one dimension with position $q$ and momentum $p$ in a
viscous environment at temperature $T$. For our first example we will consider
all continuous Markovian unravellings~\cite{WisDio01}. In an optical
realization, this includes all homodyne and heterodyne detection
schemes~\cite{WisDio01}.
The stochastic master equation (SME) describing the conditional evolution of
a particle under quantum Brownian motion (QBM), for all such
unravellings, is
\bqa
d\rho_c&=&-(i/2)\left[p^2+\left(qp+pq\right)/2,\rho_c\right]dt+ \mathcal{D}\left[ c\right]\rho_c dt \nn \\
           && +    \left\{  \sqrt{\eta}dW (c - {\rm Tr}[c\rho_{\rm c}])\rho_c + {\rm H.c.} \right\}.
  \label{qbmme}
\eqa
Here $c \equiv \sqrt{2T}q+ ip/\sqrt{8T}$ and 
\beq
  \mathcal{D}\left[B\right]A = BAB^\dagger-\mbox{$\frac{1}{2}$}\left(B^\dagger BA
                                                        - AB^\dagger B\right).
\eeq
The subscript c indicates that the evolution is conditioned on the innovation \cite{BayesBook}
 $dW$, a stochastic Wiener increment (to which we will return below). Finally, note
  that we are using scaled units such that
the damping rate $\gamma$, the particle mass $m$, and $\hbar$ are all unity.
Averaging over the noise (or setting the detection efficiency $\eta$ to zero)
removes the second line in \erf{qbmme}, leaving a version of the QBM master
equation which is in the Lindblad form~\cite{Lindblad}. It is the standard
time-independent QBM master equation~\cite{gardinerQN}, with the necessary
addition of a term which generates position diffusion~\cite{Diosi}. This system
is also equivalent to the damping of an optical cavity mode with an added
non-linearity.

The Wiener increment in the above SME satisfies $dWdW^*=dt$ and
$dW^2=\upsilon dt$, where $\upsilon= re^{i\phi} \in {\mathbb C}$ with $r \le 1$.
It is the value of $\upsilon$ which determines the measurement strategy, and thus our search
for optimally robust unravellings will involve optimization over the disk parametrized by
$r$ and $\phi$. In an optical realization, $r=0$ corresponds to heterodyne detection,
and $r=1$ corresponds to homodyne detection of a linear combination of $q$ and $p$,
$x = ce^{-i\phi/2} + c\dg e^{i\phi/2}$ ($\phi=0$ corresponds to a measurement
of $q$ and $\phi =\pi$ to one of $p$).

Since the unconditional ($\eta=0$) steady-state of Eq.~(\ref{qbmme}) possesses
a Gaussian Wigner function, and the conditional evolution preserves
Gaussianity, the four measures of robustness with which we are concerned may be
written solely in terms of the variances of $q$ and $p$ and their
covariance $C_{qp} \equiv \langle qp + pq\rangle/2 - \langle q\rangle \langle
p\rangle$.
In particular, the purity, which is required for three of the four measures, is 
$P = 1/\sqrt{4(V_qV_p - C_{qp}^2)}$.
Furthermore, these quantities evolve deterministically even under conditional
stochastic evolution. This is very useful, as it removes the need to do
stochastic simulations for the ensemble averages in Eqs.~(\ref{defPur}) --
(\ref{defSur}). This enables us to numerically optimize robustness over all the
measurement strategies (all values of $\upsilon$).


\begin{figure}
\onefigure[width=0.8\textwidth]{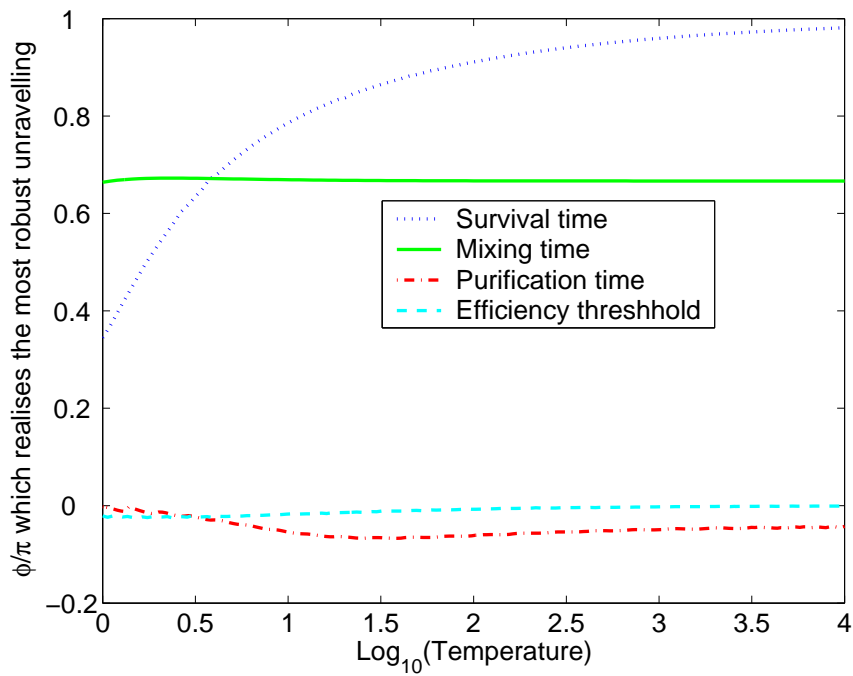}
\caption{The angle, $\phi$, which gives the optimally robust measurement scheme
for each of the four forms of robustness, plotted as a function of the
dimensionless temperature. \label{qbmsummaryfig}}
\end{figure}

Optimizing for each form of robustness over the disc of measurement strategies,
we find that the optimal strategy always lies on the boundary of the disc
($r=1$), and thus corresponds to some form of homodyne detection. However, the
type of homodyne detection required (that is, the value of the phase $\phi$) is
highly dependent upon the type of robustness desired. The results are
summarized in Figure~\ref{qbmsummaryfig}, where the optimal phase for each
notion of robustness is plotted as a function of the temperature of the bath.
We see that while the purification time and efficiency threshold may be
optimized more or less simultaneously, the mixing and survival times require
quite different unravellings \footnote{It is  worth noting a subtlety regarding
the optimal strategy for the survival time at large temperatures. While it is
true that setting $\phi=\pi$ results in a homodyne measurement of momentum, it
is incorrect to conclude from Fig.~\ref{qbmsummaryfig} that the optimal
survival time is obtained by a momentum measurement for large $T$. This is
because, for values of $\phi$ close to $\pi$, the measured observable is given
by $
   x \sim  -\frac{1}
   { \sqrt{8T}} p + (\pi-\phi) \frac{\sqrt{2T}} {2}q .
   \label{current}
$ For large $T$, the $\phi$ which maximizes $\tau_{\mbox{\scriptsize sur}}$ is
such that $(\pi-\phi)\sim T^{-1/3}$. Thus $x$ is, in fact, dominated by
position, as for the other measures.}.


\section{The two level atom}
The unconditional evolution of our second example, a driven two level atom
(TLA), is governed by the resonance fluorescence master equation \beq
   \frac{d\rho}{dt}=-i\frac{\Omega}{2}\left[ \sigma_x,\rho\right]
                                +\mathcal{D}\left[\sqrt{\gamma}\sigma_{-}\right]\rho,
\eeq where $\Omega$ is the Rabi frequency and $\gamma$ is the spontaneous
emission rate. Since the electromagnetic field is the bath which mediates the
measurement, all measurement strategies involve observing the atomic radiation.
The equation describing a particular unravelling will include additional
stochastic terms specific to the unravelling. Note that unconditional dynamics
of the TLA are entirely determined by the dimensionless parameter
$\Omega/\gamma$.

For the TLA we will consider both continuous unravellings and those which
involve discontinuous jumps. However, in this case we cannot eliminate the need
to take ensemble averages over large numbers of trajectories (for the results
which follow,  typically hundreds of thousands). As we are not therefore able
to optimize over all unravellings, we choose a small set which are  most
relevant, either from an experimental point of view, or because they posses
special properties. We consider direct photon counting, homodyne $x$ detection
($\phi=0$), homodyne $y$ detection ($\phi=\pi$), heterodyne detection, and
Adaptive Interferometric Detection (AID). This last measurement scheme was
introduced by Wiseman and Toombes~\cite{wisetoombes}, and involves interference
of the emitted radiation with a local oscillator (LO) as in homodyne detection.
However, unlike in homodyne detection, the LO is weak (comparable in amplitude
to the TLA field) so that individual photons are resolvable, resulting in a
jump process. Upon each jump the amplitude of the LO is flipped via a real-time
feedback loop, which makes it {\em adaptive} and non-Markovian. We consider AID
because it has been shown in reference~\cite{WisBra00} that it is the optimal
unravelling for maximizing the survival time, and is thus a likely candidate
for maximizing the other measures of robustness in which we are interested
here.

\begin{table}
\caption{Rankings of robustness of unravellings for each measure of robustness
with most robust at the top down to least robust at the bottom.
\label{TLAresults}}
\begin{tabular}{l|l|ll|ll} \label{Table}
$\tau_\mathrm{sur}^\mathcal{U}$ & $\tau_\mathrm{mix}^\mathcal{U}$ &
$\tau_\mathrm{pur}^\mathcal{U}$ & $\tau_\mathrm{pur}^\mathcal{U}$ &
$\eta_\mathrm{thr}^\mathcal{U}$ & $\eta_\mathrm{thr}^\mathcal{U}$ \\
&      & $\Omega\lesssim \gamma/2$ & $\Omega \gtrsim 4\gamma$ & $\Omega \simeq \gamma/2$ &
$\Omega \gtrsim 4\gamma$
\\ \hline
AID    $\,$&$\!$AID    $\,$& Hom. x $\,$&$\!\!$ Het.   $\,$&$\!\!$ AID     $\,$&$\!\!$ AID \\
Hom. x $\,$&$\!$Hom. x $\,$& AID    $\,$&$\!\!$ Hom. y $\,$&$\!\!$ Hom. x  $\,$&$\!\!$ Het.    \\
Het.   $\,$&$\!$Het.   $\,$& Direct $\,$&$\!\!$ Hom. x $\,$&$\!\!$ Direct  $\,$&$\!\!$ Hom. x  \\
Hom. y $\,$&$\!$Hom. y $\,$& Het.   $\,$&$\!\!$ AID    $\,$&$\!\!$ Het.    $\,$&$\!\!$ Hom. y  \\
Direct $\,$&$\!$Direct $\,$& Hom. y $\,$&$\!\!$ Direct $\,$&$\!\!$ Hom. y
$\,$&$\!\!$ Direct
\end{tabular}
\end{table}

We now evaluate the three measures of classical robustness for each of the
above measurement schemes, to rank them. The fourth robustness measure, the
survival time, has been calculated for this system previously
in~\cite{WisBra00}, and we include these results here for comparison. The
rankings of the unravellings are displayed in table~\ref{TLAresults}, from most to least robust.

\begin{figure}
\onefigure[width=0.8\textwidth]{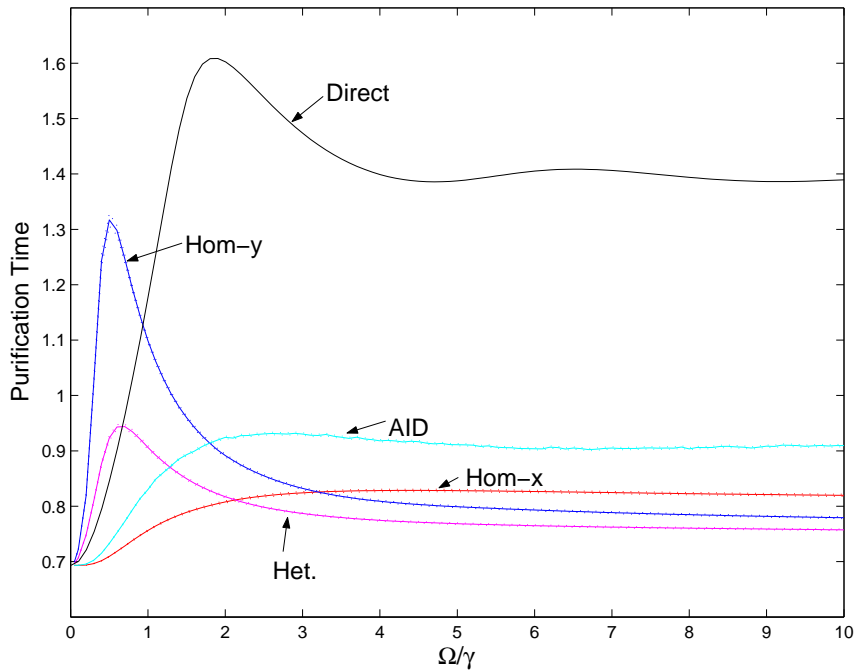}
\caption{The purification time (in units of $\gamma^{-1}$) for different measurement schemes.
\label{purtime}}
\end{figure}

The ranking of the unravellings determined by the survival and
mixing times are identical, and independent of
$\Omega/\gamma$; AID remains the most robust by these measures for all
dynamical regimes. The ranking under purification time  depends
 on the regime. For weak driving ($\Omega \lesssim \gamma/2$),
homodyne-x and AID provide the most rapid means
of obtaining information about the system. However
 for strong driving AID is one of the
least effective at extracting information. The full dependence of the
purification time for all the schemes is displayed in figure~\ref{purtime}. The
ranking in terms of the efficiency threshold also depends on the dynamical
regime as can be seen in Table \ref{Table}.  For $\gamma\ll \Omega$ (not shown
in the table) direct detection actually gives the most robust efficiency
threshold.

\section{Discussion}
We have introduced means of quantifying the degree to which quantum systems
behave in a classical fashion under continuous observation. We have also
investigated, for two canonical systems, how this classicality or robustness
depends on the way in which the environment is interrogated.

Reviewing the results for both systems reveals that for the
most part, of all the Markovian measurement schemes, homodyne detection
provides the most classically robust means of observing the systems. However,
for QBM the different concepts of robustness require {\em different} homodyne
schemes (i.e. different $\phi$). For the TLA, an adaptive (hence non-Markovian)
measurement strategy is most robust in general. Moreover, for the TLA, there
are dynamical regimes (when looking at   purification time and efficiency
threshold) where direct and heterodyne detection render the most robust
behavior. In summary, it is clear that  there is no unique unravelling which is
the most classically robust, contrary to  previous
expectations~\cite{DalDziZur01}.

The notions of robustness which we have considered here also provide an
indication of the relative merits of different measurement schemes for feedback
control constrained by time delays and measurement inefficiency. The fact that
no single unravelling is maximally robust in all ways suggests that the
measurement strategy adopted for the purposes of feedback control will need to
be tailored to the robustness requirements of a given application. Quantifying
this link between classical robustness and quantum control should provide
useful insight into the design of quantum feedback algorithms.

\acknowledgments
This work was supported by the Australian Research Council and the State of
Queensland. HMW thanks Gil Toombes for formative discussions many years ago.

\end{document}